\documentclass[%
reprint,
superscriptaddress,
noeprint,
amsmath,amssymb,
aps,
prl,
floatfix,
]{revtex4-1}

\usepackage{graphicx}
\usepackage{dcolumn}
\usepackage{bm}
\usepackage[breaklinks=true]{hyperref}
\hypersetup{colorlinks=True}

\usepackage{placeins}
\usepackage{xcolor}

\newcommand{\figscale}{0.45}

\newcommand{\thetaL}{\theta_{\mathrm{L}}}

\newcommand{\RCG}{\mathrm{RC}_{\hat{G}}}
\begin{document}


\title{Atomic nonaffinity as a predictor of plasticity in amorphous solids}

\author{Bin Xu}
\affiliation{
  Beijing Computational Science Research Center, Beijing 100193, China
}
\affiliation{
  Department of Materials Science and Engineering,  
  Johns Hopkins University, Baltimore, Maryland 21218, USA
}
\author{Michael L. Falk}
\email{mfalk@jhu.edu}
\affiliation{
  Department of Materials Science and Engineering,  Mechanical Engineering, and  Physics
  and Astronomy,
  Johns Hopkins University, Baltimore, Maryland 21218, USA
}%
\author{Sylvain Patinet}
\affiliation{
  PMMH, CNRS UMR 7636, ESPCI Paris, PSL University, Sorbonne Universit\'e, Universit\'e de Paris, F-75005 Paris, France
}
\author{Pengfei Guan}
 \email{pguan@csrc.ac.cn}
\affiliation{
  Beijing Computational Science Research Center, Beijing 100193, China
}

\date{\today}

\begin{abstract}
  Structural heterogeneity of amorphous solids present difficult challenges 
  that stymie the prediction of plastic
  events, which are intimately connected to their mechanical behavior. 
  Based on a perturbation analysis of the potential energy landscape, we derive
  the atomic nonaffinity as an indicator with intrinsic orientation, which quantifies
  the contribution of an individual atom to the total nonaffine modulus of the system.
  We find that the atomic nonaffinity can efficiently characterize the locations of 
  the shear transformation zones, with a predicative capacity comparable to 
  the best indicators. More importantly,
  the atomic nonaffinity, combining the sign of third order derivative of 
  energy with respect to coordinates, reveals an intrinsic softest shear orientation. 
  By analyzing the angle between this orientation and the shear loading direction,
  it is possible to predict the protocol-dependent response of one shear transformation zone.
  Employing the new method, the distribution of orientations of shear transformation 
  zones in model two-dimensional amorphous solids can be measured. 
  The resulting plastic events can possibly be understood from a simple model of independent 
  plastic events occurring at variously oriented shear transformation zones.
  These results shed light on the characterization and prediction of the mechanical 
  response of amorphous solids.
\end{abstract}

\maketitle




Understanding how the heterogeneity of amorphous structures correlates with
mechanical response remains a significant challenge. Various indicators have
been proposed to quantitatively predict where the material is
susceptible to plastic transformation. Some of these
only consider the structural geometry, such as free
volume~\cite{Spaepen1977,Zhu2017}, five-fold
symmetry~\cite{wakeda2007,hu2015}, and local deviation from sterically
favored structures~\cite{Tong2018}, etc. Others of these take the interaction
between particles into consideration, like low-frequency normal
modes~\cite{Maloney2004,Widmer-Cooper2008,Manning2011,Ding2014}, potential
energy~\cite{Shi2007}, local elastic modulus~\cite{Tsamados2009}, flexibility
volume~\cite{Ding2016}, mean square vibrational amplitude (MSVA)~\cite{Tong2014},
local thermal energy~\cite{Zylberg2017,Schwartzman-Nowik2019}, local yield
stress (LYS)~\cite{Patinet2016,Barbot2018}, and saddle points
sampling~\cite{Xu2018}, etc. 
Recently, machine learning has also proven to be a promising 
statistical tool to build relation between structure and plastic 
rearrangements~\cite{schoenholz_structural_2016,wang_transferable_2019,
bapst_unveiling_2020, fan_machine_2020}.

Nevertheless, most of these indicators are inherently scalar
quantities while the deformation mechanism must have an oriented shear-like 
character~\cite{Nicolas2018}. This
is clearly borne out by the fact that the orientational nature of 
shear transformation zones (STZs), the defects purported to be associated with 
plastic rearrangement, can be
measured through their high sensitivity to the deformation protocol. As
verified in simulations, under different loading orientations, 
the same glass may exhibit contrasting mechanical responses during which 
entirely different STZs are activated~\cite{Gendelman2015,Xu2018,Zylberg2017,Patinet2016,Barbot2018}.

Obtaining the mechanical response along different orientations
of one STZ may be accomplished in a number of ways: by measuring the 
LYS~\cite{Patinet2016,Barbot2018,patinet_origin_2020}, by calculating the linear response of local thermal
energy with respect to strain (LRLTE)~\cite{Schwartzman-Nowik2019}, or by sampling low-energy events~\cite{Xu2018}. 
All of these methods require computationally expensive calculations.
LYS requires prior calculations to determine the appropriate probing
length scale and direct computation of response along many orientations~\cite{Barbot2018}.
LRLTE  must be recalculated under the specified mechanical load 
to compare different orientation~\cite{Schwartzman-Nowik2019}. Sampling low-energy events 
only captures the subset of events that are inherently viscoplastic, 
and requires the harvesting of large numbers of events 
so as to find the few lowest-energy events.

In this report, based on a perturbation
analysis of the energy landscape, we derive a
parameter-free and low-cost indicator, termed 
the atomic nonaffinity.
Since this indicator is derived from a perturbation method, the atomic nonaffinity can precisely 
predict the mechanical behavior near the reference state and
and becomes less effective as the system is deformed.
We show that atomic shear nonaffinity, i.e. the shear part of the atomic 
nonaffinity, can efficiently predict 
the locations of plastic rearrangements during shear deformation
of two-dimensional amorphous solids 
with an accuracy comparable to the best known indicators. 
The relevant orientational information of STZs is naturally 
reflected in this parameter, and analysis of the atomic shear nonaffinity indicates 
that the softest shear orientation of 
the triggered STZs aligns with the orientation of the applied shear protocol.
Moreover, the distribution of orientations of activated STZs is
calculated, and we show that this distribution can be
understood through a simple model that assumes independent STZs with  
isotropically distributed soft orientations.

\begin{figure}[tb!]
  \centering
  \includegraphics[width=\figscale\textwidth]{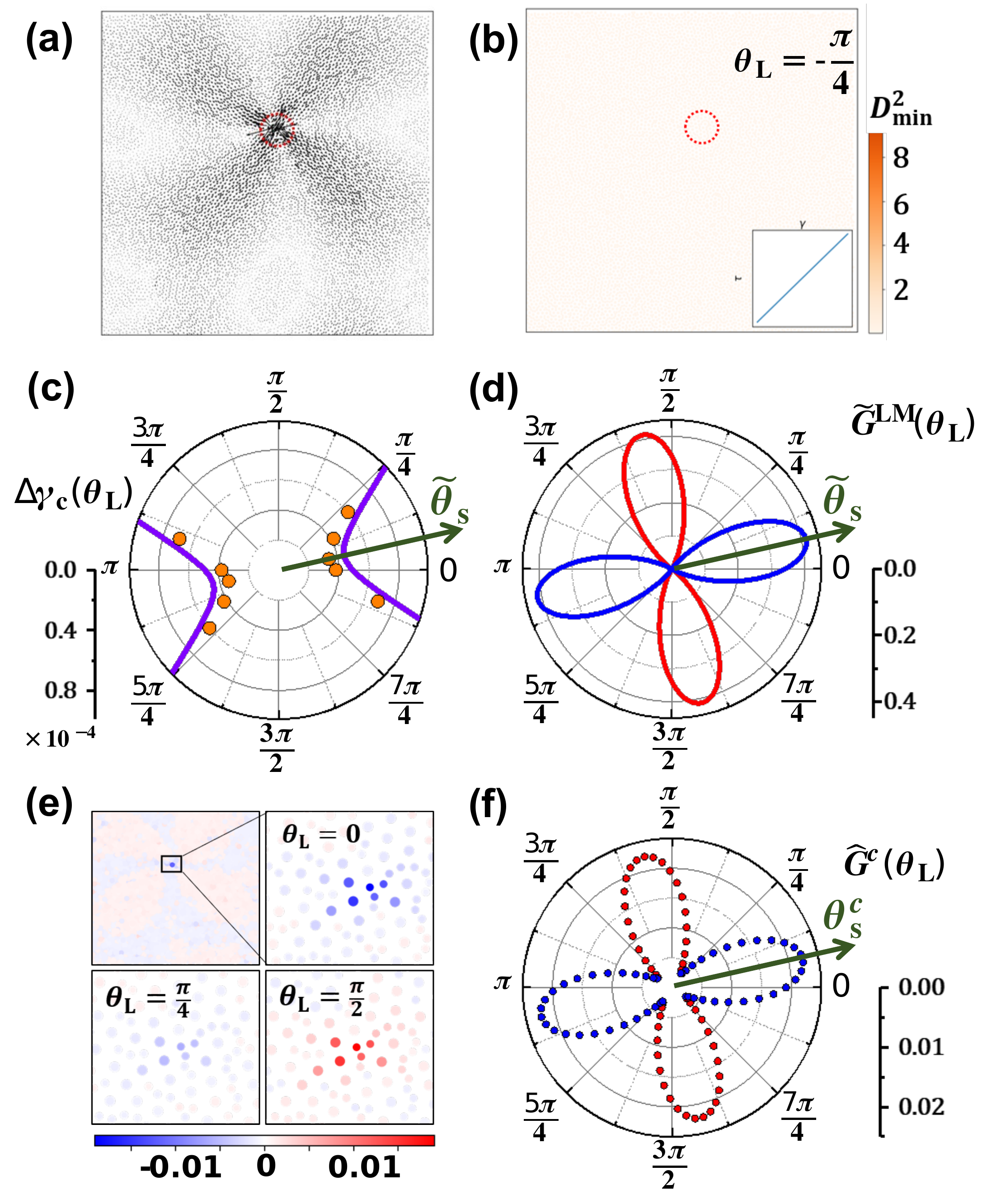}
  \caption{
    \textbf{Analysis of configuration that was sheared to be close to the triggering 
    of a plastic event.}
    (a) The spatial distribution of normal mode with lowest eigenvalue.
    (b) The $D^2_{\mathrm{min}}$ field~\cite{Falk1998} after shear with strain of 
    $\Delta\gamma=6\times10^{-5}$ in 
    the orientation of $\theta_{\mathrm{L}}=-\pi/4$. 
    The inset shows the stress-strain curve.
    (c) The predicted triggering strain (line) 
    and triggering strain from simulation (circles) as
    a function of shear angle $\theta_{\mathrm{L}}$.
    (d) The magnitude of nonaffine modulus contribution 
    from the lowest mode at different $\theta_{\mathrm{L}}$.
    Blue line represents the range of $\thetaL$, where the plastic event can be
    triggered, while the red line represents the range of $\thetaL$, where the plastic
    event can not be triggered.
    (e) The spatial distribution of $\hat{G}(\theta_\mathrm{L})\cdot \mathrm{sign}(\Delta\gamma_c(\theta_\mathrm{L}))$
    for different orientations.
    (f) The atomic shear nonaffinity in different orientations for the atom that 
    has the maximum magnitude of atomic shear nonaffinity in (e).
  }

  \label{fig:close_to_instability}
\end{figure}

~\\

To motivate the relevance of the atomic nonaffinity
we  first consider a special state in which a two-dimensional amorphous system is 
deformed to be close
to the triggering strain of a plastic event via a protocol of athermal
quasistatic shear. 
The two-dimensional glassy system, comprised of $10^4$ particles, 
was prepared via the same gradual quench
and the same smoothed Lennard-Jones potential described in 
Ref.~\cite{Barbot2018,Patinet2016}.
The spatial distribution of the normal mode with the lowest 
eigenvalue, referred to here as the lowest mode (LM), is shown
in Fig.~\ref{fig:close_to_instability}(a). A plastic event will be triggered in the region
(Fig. S1 in the supplemental materials (SM)) where the LM is localized if the system is further sheared
in this direction, denoted as the reference direction $\theta_{\text{L}}=0$. 
However, if the system is further sheared  with similar small
strains or even larger ones in other directions, such as
$\theta_{\text{L}}=-\frac{\pi}{4}$ or $\frac{\pi}{4}$, the triggering of 
the same plastic event is not observed (Fig.~\ref{fig:close_to_instability}(b) 
and Fig.~S1 in SM). 
Obviously, this protocol-dependent mechanical behavior 
of amorphous systems cannot be clearly understood solely with scalar indicators.
Here, we introduce
the second and the third derivative of the energy with respect to the vibrational
 coordinate ($q^*$) of the LM, denoted as
$\lambda^*$ and $\eta^*$ respectively, and the first derivative of 
stress of the system with respect to  $q^*$ (denoted as
$\frac{\partial\sigma_{xx}}{\partial q^*}$,
 $\frac{\partial\sigma_{yy}}{\partial q^*}$
and $\frac{\partial\tau_{xy}}{\partial q^*}$ respectively). The triggering strain
for different shear orientations can be derived as (see SM for details of derivation)
\begin{equation}
  \Delta\gamma_c(\theta_{\mathrm{L}}) =
   \frac{\lambda^{*2}}
   {2\eta^* V \frac{\partial \tau(\theta_{\mathrm{L}})}{\partial q^*}},
  \label{Equ:gammac}
\end{equation}
where $V$ is the volume of the system and $\frac{\partial
\tau(\theta_{\mathrm{L}})}{\partial q^*}$ is the first derivative of the shear stress
with respect to $q^*$ at  $\theta_{\mathrm{L}}$, which is equal to
$-\frac{1}{2}(\frac{\partial\sigma_{xx}}{\partial
q^*}-\frac{\partial\sigma_{yy}}{\partial q^*})\sin 2\theta_{\mathrm{L}}+
\frac{\partial\tau_{xy}}{\partial q^*}\cos 2\theta_{\mathrm{L}}$.
A similar form was also obtained from prior analyses of plastic mode~\cite{Lerner2016}.
Moreover, a softest shear orientation
of the LM, $\tilde{\theta}_s$, associated with the smallest positive 
triggering strain can be defined as
\begin{eqnarray}
  \label{Equ:event_theta_s}
  \tan 2\tilde{\theta}_s &= -\left( \frac{\frac{\partial \sigma_{xx}}{\partial q^*}-\frac{\partial \sigma_{yy}}{\partial q^*}}
  {\frac{2\partial\tau_{xy}}{\partial q^*}}\right), \text{with} \\ \nonumber
  \frac{\partial\tau(\theta_s)}{\partial q^*} \cdot \eta^* &> 0,
  \text{and}\ 
   \theta_s \in (-\frac{\pi}{2}, \frac{\pi}{2}].
\end{eqnarray}
Here we take the symmetry of shear into consideration and note that shear with orientation of $\theta_s$
is equal to shear with orientation of $\theta_s+\pi$. 
To verify the validity of the predictions of
Eq.~\eqref{Equ:gammac} and Eq.~\eqref{Equ:event_theta_s}, further simulations
were performed to directly measure $\Delta\gamma_c(\theta_{\text{L}})$
and $\tilde{\theta}_s$. 
As shown in Fig.~\ref{fig:close_to_instability}(c), the predictions
agree well with the simulation results, which suggests that 
the analysis of LM is successful for calculating 
the orientation-dependent mechanical response of the
system close to the instability. 
When the system is far from the instability,
we suppose that all the modes, especially the ones with small eigenvalues,
should be taken into consideration.


To develop an indicator that takes all modes into consideration while
maintaining the orientational information of each mode, we investigate
how different modes contribute to the  system modulus.
Following Maloney~\cite{Maloney2004,barron1965second},
the elastic constants of amorphous solids can be derived from the second derivative of
the total potential energy with respect to strain in athermal quasistatic deformation.
These can be rewritten in the coordinates of eigenbasis as
\begin{equation}
	C_{ijkl} = \frac{1}{V}\left( \frac{\partial^2 U}{\partial \epsilon_{ij}
	\partial \epsilon_{kl} }+ \sum_m \frac{\partial^2 U}{\partial q_m \partial \epsilon_{ij}}
\cdot \frac{dq_m}{d\epsilon_{kl}}\right),
	\label{Equ:elastic_constants}
\end{equation}
where  $U$ is the potential energy,  and $q_m$ is the $m^{\mathrm{th}}$ coordinate of
the eigenbasis of the Hessian matrix ($\frac{\partial^2
U}{\partial r_{0i}\partial r_{0j}}$).  The first term  (Born term) of
Eq.~\ref{Equ:elastic_constants}, accounts for
affine displacement and is insensitive to the structural stability~\cite{Cheng2009}  .
The second term, containing the contribution
from nonaffine relaxation in each normal mode, termed the nonaffine modulus 
($\tilde{C}$) here, is sensitive to the structural stability.
By expressing the stress as $\sigma_{ij}=\frac{1}{V}\frac{\partial
U}{\partial\epsilon_{ij}}$ and the nonaffine ``velocity'' in quasistatic
deformation  as
$\frac{dq_m}{d\epsilon_{kl}}=-\frac{1}{\lambda_m}\frac{\partial\sigma_{kl}}{\partial
q_m}$~\cite{Maloney2004}, where $\lambda_m$ is the eigenvalue of $m^{\mathrm{th}}$
normal mode, the nonaffine part, $\tilde{C}_{ijkl}$, can be rewritten as  
\begin{equation}
  \tilde{C}_{ijkl}=\sum_m \tilde{C}_{ijkl,m} = 
  \sum_m
  -\frac{V}{\lambda_m}\frac{\partial\sigma_{ij}}{\partial q_m}
  \frac{\partial\sigma_{kl}}{\partial q_m},
	\label{Equ:nonaffine_modulus}
\end{equation}
where $\tilde{C}_{ijkl,m}$ is the contribution from the $m^{th}$ normal mode, 
which is always negative.
In shear protocols, the shear modulus is the most important elastic constant. 
Thus, we focus on  the nonaffine shear modulus ($\tilde{G}$) and 
the contribution from each mode ($\tilde{G}_m$).
The $\tilde{G}_m$ can be calculated by
\begin{equation}
  \tilde{G}_{m}(\theta_{\mathrm{L}}) =  -\frac{V}{\lambda_m}
  \left( \frac{\partial\tau(\theta_{\mathrm{L}})}{\partial q_m} \right)^2,
	\label{Equ:mode}
\end{equation}
which depends on the orientation $\theta_{\mathrm{L}}$. The nonaffine shear modulus
contribution of the dominant LM, $\tilde{G}^{\mathrm{LM}}$, for the state described in 
Fig.~\ref{fig:close_to_instability}(a)-(c) is shown in 
Fig.~\ref{fig:close_to_instability}(d). The blue line 
represents the orientational range, where $\Delta\gamma_c>0$, i.e. 
$\frac{\partial\tau(\theta_{\mathrm{L}})}{\partial q^*} \cdot \eta^* >0$
following Eq.~\ref{Equ:gammac}, and the event
can be triggered. 
The red line represents the orientational range, where
the event can not be triggered.
Moreover, a softest shear orientation can also be defined 
by the largest value of $\tilde{G}^{\mathrm{LM}}$ in the blue range which is 
and should be consistent 
with the $\tilde{\theta}_s$ derived from Eq.~\ref{Equ:event_theta_s}. 

So far our results have been discussed with respect to eigenbasis. 
To develop an indicator expressed in terms of atomic quantities, we borrow an idea 
from the literature regarding the participation fraction~\cite{Widmer-Cooper2008,Ding2014}.
By expressing the normalized eigenvector in the atomic coordinates   
as $\bm{\Psi}_m = \sum_{n,\alpha}
c_{mn\alpha}\bm{e}_{n\alpha}$, where $\bm{e}_{n\alpha}$ is a unit vector
corresponding to the displacement of $n^{th}$ atom in the $\alpha (=x
\, \mathrm{or}\, y)$ direction, and $c_{mn\alpha}$ is the projection of the
$m^{th}$ eigenvector onto  $\bm{e}_{n\alpha}$, the $\tilde{C}_{ijkl}$
can be rewritten as
\begin{equation}
  \tilde{C}_{ijkl}=\sum_n\hat{C}_{ijkl,n} =\sum_n \sum_{m,\alpha}-\frac{V}{\lambda_m}
  \frac{\partial \sigma_{ij}}{\partial q_m}
  \frac{\partial \sigma_{kl}}{\partial q_m}
	c_{mn\alpha}^2.
	\label{Equ:atomic_nonaffine_modulus}
\end{equation}
Here $\hat{C}_{ijkl,n}$ is the atomic nonaffinity of the $n^{th}$ atom.
As most local plastic
rearrangements are shear-like~\cite{Argon1979,Falk1998}, the atomic 
shear nonaffinity (ASN) is the most important component when investigating the STZs and 
can be written as
\begin{equation}
  \hat{G}_{n}(\theta_{\mathrm{L}}) = \sum_{m,\alpha}-\frac{V}{\lambda_m}
  \left( \frac{\partial \tau(\theta_{\mathrm{L}})}{\partial q_m} \right)^2 c_{mn\alpha}^2.
  \label{Equ:shear_nonaffinity}
\end{equation}
Obviously, the value of $\hat{G}_n$ depends on the orientation $\thetaL$.
As a result, the spatial distribution of $\hat{G}_n$ in the previous case shown 
in Fig.~\ref{fig:close_to_instability}(e) exhibits a clear orientation-dependent 
behavior in the region where the plastic event is located.
More negative values of $\hat{G}_n$ mean that the corresponding atom is 
easier to trigger in the orientation $\thetaL$. The $\hat{G}_n$
distribution calculated at different orientations indicates that $\thetaL=0$ is the easiest shear direction 
for the plastic event when compared with $\thetaL=\frac{\pi}{4}$ and $\frac{\pi}{2}$, 
which is consistent with the direct loading 
test in Fig.~\ref{fig:close_to_instability}(b) and Fig.~S1 in the SM.
Moreover, the atom located in the core region has the maximum magnitude of 
atomic shear nonaffinity, denoted as $\hat{G}^c(\thetaL)$.
Fig.~\ref{fig:close_to_instability}(f) shows the $\thetaL$-dependent $\hat{G}^c$,
which has a similar shape as the $\tilde{G}^{\mathrm{LM}}$. 
This is attributed to the fact that the LM with 
smallest eigenvalue dominates the variation of $\hat{G}^c$,  which
can be inferred from Eq.~\ref{Equ:shear_nonaffinity}. 
Thus, we can define the softest shear orientation for the
$n^{th}$ atom as the softest shear orientation of the mode that dominates the variation
of $\hat{G}_n$. The softest shear orientation of the
$n^{\mathrm{th}}$ atom is defined as
\begin{equation}
  \theta_{n,s} = \tilde{\theta}_{i,s}, i = \mathrm{argmax}_m \sum_{\alpha} |\tilde{G}_m(\tilde{\theta}_{m,s})| c_{mn\alpha}^2,
  \label{Equ:theta_atom}
\end{equation}
and the calculated softest shear orientation of the core atom  ($\theta_s^c$) 
is shown in Fig.~\ref{fig:close_to_instability}(f). 
The consistency of the proposed softest shear orientations for 
one STZ from the three parameters, i.e. the directly calculated triggering 
strain (Fig.~\ref{fig:close_to_instability}(c)), 
nonaffine modulus of the lowest mode (Fig.~\ref{fig:close_to_instability}(d)),
and the atomic shear nonaffinity (Fig.~\ref{fig:close_to_instability}(f)), 
implies that the $\theta_s$ defined 
from atomic shear nonaffinity is effective to for characterizing  the 
orientations of STZs.

\begin{figure}[tb!]
  \centering
  \includegraphics[width=\figscale\textwidth]{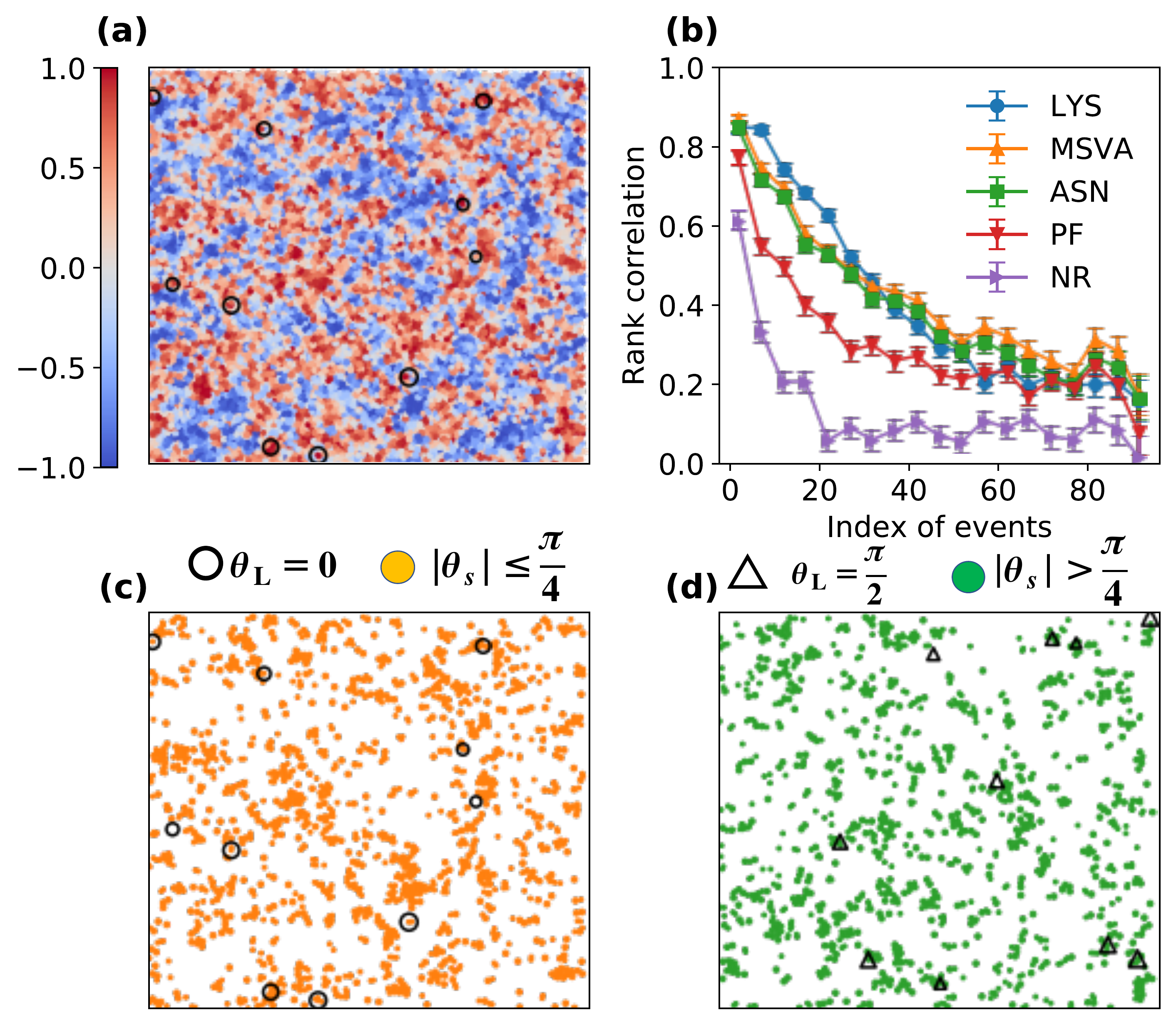}
  \caption{
  \textbf{Predicting plastic events by analyzing initial configurations.}
  (a) The correlation field of atomic shear 
  nonaffinity $\hat{G}(\theta_{\mathrm{L}}=0)$ and 
  the locations of the first ten plastic events (black circles) triggered 
  in shear protocols with $\theta_{\mathrm{L}}=0$.
  (b) Correlation between the indicators including local yield stress (LYS), mean square vibrational amplitude (MSVA),
  atomic shear nonaffinity (ASN), participation fraction (PF), nonaffine rate (NR) with the locations
  of plastic events as a function of the index of the events. Averages are taken over windows of five
  events. Error bar at each window represents the standard deviation of the mean.
  (c) Orange circles represent the atoms 
  with $\RCG>0$ and $|\theta_s|< \frac{\pi}{4}$.
  Black circles mark the locations of the first ten plastic events with $\thetaL=0$.
  (d) Green circles represent the atoms 
  with $\RCG>0$ and $|\theta_s|> \frac{\pi}{4}$.
  Triangles mark the locations of the first ten plastic events 
  with $\thetaL= \frac{\pi}{2}$.
}
  \label{fig:correlation_and_distribution}
\end{figure}

~\\
Now that we have seen the predictive capacity of atomic shear nonaffinity regarding the 
protocol-dependent mechanical response of  a plastic event close to 
instability we can ask, "What if the system is not close to  instability?"
and, "How predictive is this indicator?"
Predicting plastic events in an amorphous system by analyzing
the local indicators of initial structure has been extensively 
studied in the 
literature~\cite{Patinet2016,Ding2014,Manning2011,Xu2018,Zylberg2017,Widmer-Cooper2008,Cubuk2015,Lerner2016,Gendelman2015,richard_predicting_2020}.
To compare the reliability of local indicators for predicting plastic 
events, one hundred two-dimensional samples 
prepared with the same thermal history as the previous sample 
were employed for local properties 
calculations. 
The athermal quasistatic 
shear deformation with a strain step of $\Delta\gamma_{xy}=10^{-5}$ was 
then applied to each sample, and each stress drop in the stress-strain curve 
was associated with the resulting atomic rearrangements corresponding to one 
plastic event. Nonaffine rate~\cite{Maloney2004}
was calculated for the configurations just before the stress drops, and 
the atom with maximum nonaffine rate 
was identified as the core atom, whose index is denoted as $\mathrm{ID}_N$ for
$N^{th}$ plastic event.
To compare the success of different indicators, we transform
those indicators to a rank correlation (RC) value following the analysis performed 
by Patinet et al.~\cite{Patinet2016} as
\begin{equation}
  \mathrm{RC}_{\Psi}(n) = 1 - 2 \mathrm{CDF}_{\Psi}(n),
	\label{Equ:correlation}
\end{equation}
where $\Psi$ is one of the indicators,
$\mathrm{CDF}_{\Psi}$ is the cumulative distribution function for the $\Psi$ of all
atoms, and $\mathrm{CDF}_{\Psi}(n)$ is the function value in the 
range of $[0,1]$ based on the value of 
$\Psi$ for $n^{th}$ particle.
The spatial distribution field of the calculated $\mathrm{RC}_{\hat{G}}$ with 
$\thetaL=0$ is shown in Fig.~\ref{fig:correlation_and_distribution}(a).
The first ten plastic events in shear protocols with $\thetaL=0$ 
are almost all located at high $\mathrm{RC}_{\hat{G}}$ regions, 
which implies the highly predictive 
power of $\hat{G}$. To quantitatively compare the predictive power regarding
plastic events for different local indicators, the relationship between the 
locations of plastic events 
and the corresponding values of local indicators is described 
by the average of $\mathrm{RC}_{\Psi}(\mathrm{ID}_N)$ 
over 100 samples. The average $\mathrm{RC}_{\Psi}(\mathrm{ID}_N)$ of investigated local 
indicators , such as 
participation fraction (PF)~\cite{Manning2011,Ding2014,Widmer-Cooper2008} in the lowest 
$1\%$ of normal modes,
the nonaffine rate (NR)~\cite{Maloney2004},
the MSVA~\cite{Tong2014}, the LYS~\cite{Patinet2016} and the ASN, are shown in 
Fig.~\ref{fig:correlation_and_distribution}(b).
The LYS presents the highest predictive power in the early stage, since  
nonlinear response to shear is considered. 
The  MSVA, and ASN shows comparable predictive power, 
 and the other indicators
have lower predictive power than those three. 
It is worth noting that the predicative power of the indicators 
depends on the stability of configurations. In the SM we show 
the predictive power of these indicators for configurations 
prepared by instantly quenching
from high temperature liquids, systems in which MSVA and
ASN outperform LYS.
We also note that the
orientational information in ASN is incomplete and it, as a modulus, has 
the same value for $\thetaL=0$ and $\frac{\pi}{2}$, while local regions 
generally have different mechanical behavior for those two protocols.


~\\
As discussed in Fig.~\ref{fig:close_to_instability}, 
we expect that the plastic events induced when
shearing along direction $\thetaL$ should be located at the atoms  with 
$|\theta_{s}-\thetaL|<\frac{\pi}{4}$, and here we test this 
expectation in one of the previous samples.
We focused on the "soft" atoms in the sample with 
$\mathrm{RC}_{\hat{G}}>0$, and distinguished them by the value of $\theta_s$.
The correlation between the atom with $\RCG>0$ and
$|\theta_s|< \frac{\pi}{4}$ ($|\theta_s|>\frac{\pi}{4}$) and the first ten 
plastic events of $\thetaL=0$ ($\thetaL=\frac{\pi}{2}$) direction is 
illustrated in 
Fig.~\ref{fig:correlation_and_distribution}(c) 
(Fig.~\ref{fig:correlation_and_distribution}(d)).
The correlation in both 
Fig.~\ref{fig:correlation_and_distribution}(c) and (d) indicate 
that the predictive power can potentially be increased by 
screening for regions where the intrinsic softest 
orientation of $\hat{G}_n$ aligns with the deformation protocol. (Similar results about protocols of 
$\thetaL=\frac{\pi}{4}$ and $-\frac{\pi}{4}$ are shown in SM.)

\begin{figure}[tb!]
  \centering
  \includegraphics[width=\figscale\textwidth]{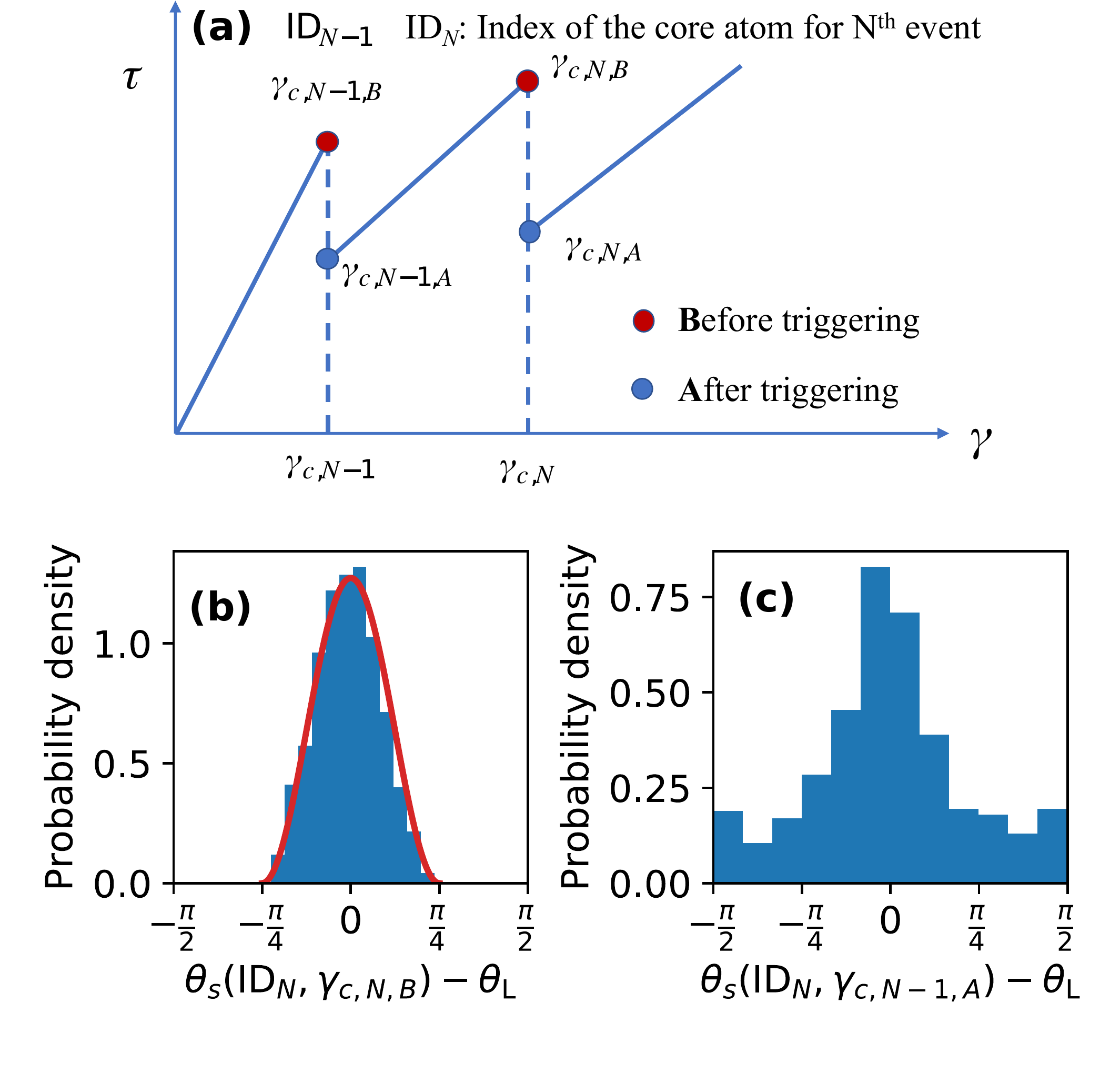}
  \caption{
  \textbf{Ditribution of the softest shear orientations for plastic regions.}
  (a) A schematic diagram introducing the notations used in this figure.
  (b)  The distribution of the softest shear orientations for all the plastic events
  that are triggered before shear strain $0.12$ in 10 samples. The softest shear orientations
  are calculated based on the configuration just before each event.
  Red line follows the function $\frac{4}{\pi}\cos^2(2\Delta\theta)$.
  (c)  Similar to (b), but the softest shear orientations for each event are calculated
  based on the configurations  just after the previous event. 
}
  \label{fig:orientations}
\end{figure}

However, there still exist some number of events
that are not caught by the criterion $|\theta_{s}-\thetaL|< \frac{\pi}{4}$.
This can be attributed to the rotation of $\theta_s$ during deformation,
since the $\theta_s$ is calculated mainly based on 
a second-order perturbation method, and higher order terms and nonlinear interactions
between different modes can lead to the rotation of $\theta_s$.
To obtain the statistics of the rotation of $\theta_s$,
the softest shear orientations of core particles 
of all the plastic events before shear strain $0.12$ with $\thetaL=0$
in 10 samples are calculated
based on the configurations just before each event or just after the last 
event (illustrated in Fig.\ref{fig:orientations}(a)).
The distribution of the calculated $\theta_s$ of those core particles 
in the configurations just before triggering
are shown in Fig.~\ref{fig:orientations}(b) and all $\theta_s$ satisfy 
the criterion $|\theta_s-\theta_{\mathrm{L}}|<\frac{\pi}{4}$, which 
is what we expected for systems close to instability, as discussed previously.
Moreover, the peak of probability density is located at
$\theta_{\mathrm{L}}$ implying that the region with the intrinsically softest 
orientation closest to the imposed shear orientation
is easiest to trigger.
However, the distribution is broadened as shown in 
Fig.~\ref{fig:orientations}(c) for the calculated $\theta_{s}^c$
based on the configurations just after the triggering of the previous
event. In this analysis only approximately $75\%$ of the plastic events 
satisfy the criterion.
More statistics about how orientations calculated 
by our perturbation method change are presented in SM.
Because plastic events tend to happen at STZs 
closely aligned with orientation of the shear protocols, we also show that 
that the precision of prediction for 
different indicators can be improved by screening for potential
 STZs with the softest shear orientations in SM.

The distribution of orientations of the triggered plastic event shown in 
Fig.~\ref{fig:orientations}(b) is regular.
It can be understood 
by a simple model of independent 
plastic events with intrinsic orientations. 
In this model, we assume that amorphous solids are
isotropic and that the shear-orientation-dependent 
triggering strain can be derived from Eq.~\ref{Equ:gammac} as
\begin{equation}
  \gamma_c(\theta_{\mathrm{L}}) = \frac{\gamma_c(\theta_s)}
  {\cos[2(\theta_s-\theta_{\mathrm{L}})]},
  \label{Equ:gammac_theta}
\end{equation}
where $\theta_s$ is the softest shear orientation of a STZ. 
If we assume that the 
number density for a particular softest shear direction $\theta_s$ at different triggering
strains $\gamma_c(\theta_s)$ (noted as $\gamma_{c,s}$) follows a power law 
$\rho(\gamma_{c,s})=A\gamma_{c,s}^\alpha$~\cite{Karmakar2010c,Hentschel2015,Lin14382,Lin_2014}, the probability density
distribution of orientations 
$\hat{\rho}(\theta_s-\theta_{\mathrm{L}})$ 
(denoted as $\hat{\rho}(\Delta\theta)$) will follow  
(see SM for details of derivation)
\begin{equation}
  \hat{\rho}(\Delta\theta) = k \cos^{\alpha+1} (2\Delta\theta)
  \label{Equ:ratio}
\end{equation}
The probability density distribution in Fig.~\ref{fig:orientations}(b) 
corresponds to $\alpha=1$, as it closely fits
a distribution function $\frac{4}{\pi}\cos^2(2\Delta\theta)$ (red line in 
Fig.~\ref{fig:orientations}(b)).
These results are also supported
by the  probability distribution function of local yield stress
of the samples, in which $\alpha \approx 1.1$, as
shown in Ref.\cite{Patinet2016}.

~\\
In summary,
we have derived a general and parameter-free indicator, the
atomic nonaffinity. It is well-defined and is easy to apply in systems
beyond the 2d Lennard-Jones system discussed here. The atomic nonaffinity has a
clear physical meaning in that the summation of atomic nonaffinities
corresponds to the total nonaffine modulus of the system.  The softest shear
orientation of each region  is defined based on the atomic shear nonaffinity
and stems from anisotropy of the shear stress derivative against the coordinate of
the low-frequency mode in different orientations. When combined with the sign 
of the third order derivative of energy with respect to coordinates, it reveals the intrinsic
orientation of the  plastic rearrangement  and 
directly connects to the anisotropic mechanical response of local regions,
which is important for understanding aspects of the mechanical behavior of amorphous
solids  not directly reflected or defined in other indicators.  As atomic
nonaffinity is developed based on the nonaffine response of atoms upon
deformation, it naturally has a good correlation with the plastic events,
comparable to the best indicators.
Mechanical behavior must be correlated with structure,
and we anticipate that this method will be important for elucidating the
structural origin of the anisotropic mechanical response in specific systems.

%
%
 
\begin{acknowledgements}
B.X. and P.F.G acknowledge financial support by the 
National Natural Science Foundation of China
(NSFC, Grants No.~51571011/U1930402), 
the MOST 973 program (No.~2015CB856800).  
M.L.F. acknowledges support provided by NSF Grant Award No.~1910066/1909733. 
We acknowledge the computational support from 
the Beijing Computational Science Research Center.
\end{acknowledgements}

\FloatBarrier


%

%
%

\end{document}